\documentclass[10pt,twocolumn,a4paper,aps,prd,
preprintnumbers,showpacs,superscriptaddress,
nofootinbib,amsmath,amssymb
]{revtex4-1}
\usepackage{graphicx}
\usepackage{hyperref}
\usepackage{subfigure}
\usepackage{multirow}
\usepackage[toc,page]{appendix}
\usepackage{ulem}
\usepackage{adjustbox}
\DeclareMathAlphabet{\pazocal}{OMS}{zplm}{m}{n}

\begin{document}
\title{Quantum corrected black holes: quasinormal modes, scattering, shadows}
\author{R. A. Konoplya}\email{roman.konoplya@gmail.com}
\affiliation{Institute of Physics and Research Centre of Theoretical Physics and Astrophysics, Faculty of Philosophy and Science, Silesian University in Opava, CZ-746 01 Opava, Czech Republic}
\affiliation{Peoples Friendship University of Russia (RUDN University), 6 Miklukho-Maklaya Street, Moscow 117198, Russian Federation}
\begin{abstract}
The spherically symmetric deformation of the Schwarzschild solution owing to the quantum corrections to gravity is known as Kazakov-Solodukhin black-hole metric. Neglecting non-spherical deformations of the background the problem was solved non-perturbatively. Here we analyze the basic characteristics of this geometry, such as: quasinormal modes and grey-body factors of fields of various spin and shadow cast by this black hole. The WKB approach and time-domain integration method, which we used for calculation of quasinormal modes, are in a good concordance. The analytical formula for quasinormal modes is deduced in the eikonal regime. The radius of shadow is decreasing when the quantum deformation is turned on.
 \end{abstract}
\maketitle

\section{Introduction}\label{introduction}

According to various approaches to quantization of gravity the Schwarzschild solution representing a spherically symmetric black hole in General Relativity must acquire quantum corrections. Some approaches for finding such quantum corrected black-hole geometry are related to the semi-classical perturbative corrections owing to the polarization of vacuum by matter fields in the vicinity of a black hole \cite{YorkPage}. The consequent analysis of the proper oscillations frequencies of such black holes was done in \cite{Piedra:2009pf,Piedra:2010ur}. However, this approach does not take into consideration contribution of, apparently, the dominating factor - quantization of the gravitational fields itself. Fortunately, when neglecting the non-spherical deformations and using the effective scalar-tensor gravity, the problem proved to be renormalizable and the corresponding generalization of the Schwarzschild solution was found long time ago by Kazakov and Solodukhin \cite{Kazakov:1993ha}. Thermodynamic properties of these black holes were considered in  \cite{Kim:2012cma,Shahjalal:2019ypz,Lobo:2019put}, and some cosmological applications in \cite{Shahjalal:2019fig}.

Here we would like to consider the basic characteristics related to potentially observable quantities of the Kazakov-Solodukhin black hole. One of such characteristics of the black-hole geometry is the set of proper oscillation frequencies (termed  ``quasinormal modes'') governing the relaxation of perturbations in the vicinity of a black hole at intermediately late times. These frequencies are called ``the fingerprints'' of a black hole, because they do not depend on the way of excitation and characterize the black-hole geometry itself. The other characteristic is the grey-body factors which say how much of the initial flux of Hawking radiation will penetrate the potential barrier and reach a distant observer. Finally, the shadows cast by a black hole is important to determine the near horizon geometry (see for example \cite{Bisnovatyi-Kogan:2017kii,Younsi:2016azx,Konoplya:2019sns,Dokuchaev:2019jqq,Tian:2019yhn,Javed:2019rrg,Zhu:2019ura,Contreras:2019nih,Long:2019nox} and references therein).
The current experiments related to detection of gravitational waves from black holes as well as observations of the electromagnetic spectrum in their vicinity
\cite{alternative}  leave an ample room for interpretation of the near horizon geometry of black holes, because the parameters of black holes and the models for the matter, surrounding them, are known with large uncertainty and freedom of interpretation.

One of the above three characteristics, quasinormal spectrum, was previously considered in \cite{Saleh:2016pke,Saleh:2014uca}. However, the data presented in \cite{Saleh:2016pke,Saleh:2014uca} is, in our opinion, not complete and fully satisfactory. First of all, there were presented quasinormal modes computed by the third order WKB formula in the regime when the overtone number $n$ is larger than the multipole number $\ell$, that is, when the WKB method does not work. Then, the lowest (and most important) multipoles $\ell$ are not presented there for the scalar and Dirac fields, apparently because the third order WKB formula is not accurate and lead to meaningless results in this case. Finally, no electromagnetic modes and data for the near extremal values of the deformation parameter were presented. The gravitational perturbations considered in \cite{Saleh:2016pke,Saleh:2014uca} were treated wrongly, because the equation $\delta R_{\mu \nu} =0$ was used for the spacetime which is not Ricci-flat. Here we will complement and correct these results by considering quasinormal spectrum of test scalar, Dirac and electromagnetic fields and a conformally coupled scalar field. In addition, we will consider the scattering of fields and show the dependence of the radius of the black hole shadow on the deformation parameter.

The paper is organized as follows. The Sec. II is devoted to the essentials of the Kazakov-Solodukhin metric. Sec. III gives the detailed analysis of quasinormal modes of minimally coupled scalar, electromagnetic and Dirac fields via higher order WKB approach combined with the Pade approximation and by the time-domain integration. Sec. IV we summarize the results of calculations of quasinormal modes for the conformally coupled scalar field. In Sec. V we solve the scattering problem. In Sec. V we find the radius of the shadow cast by the quantum corrected black hole.

\section{Kazakov-Solodukhin metric}

The deformation of the Schwarzschild solution in general relativity due to spherically symmetric quantum fluctuations of the metric was obtained by Kazakov and Solodukhin in \cite{Kazakov:1993ha}. In that case, the 4D theory of gravity with Einstein action reduces to the
effective two-dimensional dilaton gravity.
The Kazakov-Solodukhin metric \cite{Kazakov:1993ha} has the form
\begin{equation}
d s^2 = - f(r) d t^2 + f^{-1}(r) d r^2 + r^2 (d \theta^2 + sin^2 \theta d \varphi^2).
\end{equation}
where the renormalizable potential  $U(\rho)$ has the following form
\begin{equation}\label{4}
    U(\rho)=\frac{e^{-\rho}}{\sqrt{e^{-2\rho}-\frac{4}{\pi}G_R}},
\end{equation}
where $G_R=G_Nln(\mu/\mu_0)$, $G_N$ is the Newton constant and $\mu$ is a scale parameter.
Then, the metric function $f(r)$ of the quantum-corrected Schwarzschild black hole is
\begin{equation}\label{2}
    f(r)=-\frac{2M}{r}+\frac{1}{r}\int^r U(\rho)d\rho,
\end{equation}
which gives
\begin{equation}
f(r)=\left(\frac{\sqrt{r^2-a^2}}{r}-\frac{2 M}{r}\right).
\end{equation}
Here $a^2=4G_R/\pi$ is the deformation parameter.
For an empty space, $U(\rho)=1$ and the metric is reduced to the Schwarzschild one. The spacetime is not Ricci-flat  and the curvature is
\begin{equation}\label{eq:R}
R(r) = \frac{2}{r^2} \left(1-\frac{1}{\sqrt{1-\frac{a^2}{r^2}}}\right)+\frac{a^2}{r^4} \left(1-\frac{a^2}{r^2} \right)^{-\frac{3}{2}}. 
\end{equation}
The event horizon $r_h$ is situated at $r= r_h =\sqrt{4 M^2 + a^2}$.
The singularity is located at $r =a$, because in the limit $a \rightarrow r$ the curvature goes to infinity.

\section{Quasinormal modes}

\begin{table}
\begin{tabular}{p{1cm}cccc}
\hline
$a$ & WKB (7th order, $\tilde{m} =6$) & Time-domain  \\
\hline
0.01   & $0.223624-0.210830 i$ & $0.22112 - 0.21057 i$  \\
0.1    & $0.223962-0.211724 i$ &  $0.22147 - 0.21144 i$  \\
0.2    & $0.224992-0.214503 i$ &  $0.22237 - 0.21420 i$  \\
0.3    & $0.226730-0.219381 i$ &   $0.22437 - 0.21867 i$  \\
0.4    & $0.229239-0.226810 i$ &  $0.22690 - 0.22580 i$  \\
0.5    & $0.232801-0.237722 i$ &$0.23012 - 0.23620 i$  \\
0.6    & $0.238824-0.254580 i$  & $0.23373 - 0.25193 i$  \\
0.7    & $0.255035-0.287220 i$ &  $--$ \\
0.8    & $0.355043-0.411972 i$ &  $--$  \\
0.9    & $0.492339-0.102714 i$ & $--$  \\
\hline
\end{tabular}
\caption{The fundamental quasinormal mode of the scalar field ($\ell=0$, $n=0$, $r_{h} =1$) for various values of $a$. The scalar quasinormal frequencies for $\ell=0$ cannot be easily extracted from the time domain profile at large $a$, because the asymptotic tail dominates after only a few oscillations, which is not enough for the Prony method. }\label{tab0}
\end{table}

\begin{table*}
\begin{tabular}{p{1cm}cccc}
\hline
$a$ & WKB for $V_{+1/2}$, $V_{-1/2}$ (7th order, $\tilde{m} =6$)  & Time-domain  for $V_{-1/2}$  \\
\hline
0.01    & $0.365864-0.193605 i$, $0.365885-0.194079 i$ & $0.36577 - 0.19444 i$ \\
0.1    & $0.366368-0.194364 i$, $0.366386-0.194844 i$ &  $0.36627 - 0.19415 i$ \\
0.2    & $0.367919-0.196723 i$, $0.367934-0.197223 i$ &  $0.36792 - 0.19632 i$ \\
0.3   & $ 0.370599-0.200864 i$, $0.370607-0.201403 i$ &   $0.37052 - 0.20047 i$  \\
0.4    & $0.374567-0.207155 i$, $0.374564-0.207768 i$ &  $0.37448 - 0.20683 i$ \\
0.5      & $0.380095-0.216257 i$, $0.380077-0.217022 i$ & $0.38011 - 0.21594 i$  \\
0.6    & $0.387673-0.229352 i$, $0.387768-0.230626 i$  & $0.38735 - 0.22921 i$  \\
0.7    & $0.398011-0.248958 i$, $0.399390-0.251245 i$ &  $0.39722 - 0.24915 i$ \\
0.8    & $0.412694-0.280829 i$, $0.418123-0.283572 i$ &  $0.41127 - 0.28165 i$  \\
0.9    & $0.340357-0.384698 i$, $0.462018-0.350993 i$ & $0.43288 - 0.35007 i$ \\
\hline
\end{tabular}
\caption{The fundamental quasinormal mode of the Dirac field ($k=1$, $n=0$, $r_{h} =1$) for various values of $a$.}\label{tab1}
\end{table*}

\begin{table}
\begin{tabular}{p{1cm}cccc}
\hline
$a$ & WKB (7th order, $\tilde{m} =6$)   & Time-domain  \\
\hline
0.01    & $0.496530-0.184981 i$ & $0.49656 - 0.18497 i$  \\
0.1    & $0.497119-0.185655 i$ &  $0.49727 - 0.18553 i$  \\
0.2    & $0.498927-0.187745 i$ &  $0.49907 - 0.18764 i$  \\
0.3   & $0.502026-0.191398 i$ &   $0.50215 - 0.19131 i$  \\
0.4    & $0.506553-0.196905 i$ &  $0.50666 - 0.19684 i$  \\
0.5      & $0.512725-0.204767 i$ &$0.51281 - 0.20475 i$  \\
0.6    & $0.520870-0.215835 i$  & $0.52096 - 0.21595 i$  \\
0.7    & $0.531267-0.231269 i$ &  $0.53162 - 0.23214 i$ \\
0.8    & $0.537418-0.257872 i$ &  $0.54539 - 0.25698 i$  \\
0.9    & $0.543220-0.314854 i$ & $0.56216 - 0.30067 i$  \\
\hline
\end{tabular}
\caption{The fundamental quasinormal mode of the electromagnetic field ($\ell=1$, $n=0$, $r_{h} =1$) for various values of $a$.}\label{tab2}
\end{table}

\begin{table}
\begin{tabular}{p{1cm}cc}
  \hline
a&   $\omega $ (Time-domain)\\
 \hline
0.9&  $0.56216 - 0.30067 i$\\
0.91& $0.56383 -0.30706 i$\\
0.92& $0.56543 - 0.31405 i$\\
0.93& $0.56691 - 0.32174 i$\\
0.94& $0.56821 - 0.33026 i$\\
0.95& $0.56925 - 0.33995 i$\\
0.96& $0.56985 - 0.35064 i$\\
0.97& $0.56981 - 0.36316 i$\\
0.98& $0.56871 - 0.37814 i$\\
0.99& $0.56573 - 0.39746 i$\\
0.992& $0.56452 - 0.40216 i$\\
0.994& $0.56318 - 0.40751 i$\\
0.996& $0.56145 - 0.41373 i$\\
0.998& $0.55899 - 0.42160 i$\\
0.999& $0.55714 - 0.42699 i$\\
  \hline
\end{tabular}
\caption{The fundamental quasinormal mode of the electromagnetic field ($\ell=1$, $n=0$, $r_{h} =1$) for the near extremal black hole.}\label{tab3}
\end{table}

The general covariant equation for a massless scalar field has the form
\begin{equation}\label{KGg}
\frac{1}{\sqrt{-g}}\partial_\mu \left(\sqrt{-g}g^{\mu \nu}\partial_\nu\Phi\right)=0,
\end{equation}
and for an electromagnetic field it has the form
\begin{equation}\label{EmagEq}
\frac{1}{\sqrt{-g}}\partial_\mu \left(F_{\rho\sigma}g^{\rho \nu}g^{\sigma \mu}\sqrt{-g}\right)=0\,,
\end{equation}
where $F_{\rho\sigma}=\partial_\rho A_{\sigma}-\partial_\sigma A_{\rho}$ and $A_\mu$ is a vector potential.

For the general background spacetime, the massless Dirac equation is
\begin{equation}\label{dirac}
    [\gamma^ae^\mu_a(\partial_\mu+\Gamma_\mu)]\Psi=0,
\end{equation}
where $\gamma^a$ is the Dirac matrix, $e^\mu_a$ is the inverse of the tetrad $e^a_\mu$ ($g_{\mu\nu}=\eta_{ab}e^a_\mu e^b_\nu$), $\eta_{ab}$ is the Minkowski metric.
The spin connections $\Gamma_\mu$ are defined as follows
\begin{equation}
\Gamma_\mu=\frac{1}{8}[\gamma^a, \gamma^b]e^\nu_ae_{b\nu;\mu}.
\end{equation}

After separation of the variables Eqs. (\ref{KGg}), (\ref{EmagEq}) and (\ref{dirac}) take the following general wave-like form
\begin{equation}\label{wave-equation}
\frac{d^2\Psi_s}{dr_*^2}+\left(\omega^2-V_{s}(r)\right)\Psi_s=0,
\end{equation}
where $s=0$ corresponds to scalar field, $s=\pm1/2$ to Dirac field, and $s=1$ to the electromagnetic field. The "tortoise coordinate" $r_*$ is defined by the relation
$ dr_*=dr/f(r)$, and the effective potentials are
\begin{equation}\label{scalarpotential}
V_{0}(r) = f(r)\left(\frac{\ell(\ell+1)}{r^2}+\frac{1}{r}\frac{d f(r)}{dr}\right),
\end{equation}
\begin{equation}\label{empotential}
V_{1}(r) = f(r)\frac{\ell(\ell+1)}{r^2}.
\end{equation}
\begin{equation}
       V_{\pm 1/2} = \frac{\sqrt{f}|k|}{r^2}\Big(|k|\sqrt{f} \pm \frac{r}{2}\frac{df}{dr} \mp f\Big)\\
\end{equation}
where $|k| = 1, 2, 3, \cdot$ is the total angular momentum number \cite{Brill:1957fx}. As can be seen from \cite{Zinhailo:2019rwd} for generic spherically symmetric spacetimes, both effective potential $V_{+1/2}$ and $V_{-1/2}$  are iso-spectral.

Quasinormal modes $\omega_{n}$ correspond to solutions of the master wave equation (\ref{wave-equation}) with the requirement of the purely outgoing waves at infinity and purely incoming waves at the event horizon:
\begin{equation}
\Psi_{s} \sim \pm e^{\pm i \omega r^{*}}, \quad r^{*} \rightarrow \pm \infty.
\end{equation}

In order to find quasinormal modes we shall use two independent methods:
\begin{enumerate}
\item the integration of the wave equation (before introduction the stationary ansatz, that is, with the second derivative in time instead of $\omega^2$-term) in time domain at a given point in space \cite{Gundlach:1993tp}
\item the WKB method suggested by Will and Schutz \cite{Schutz:1985zz}, extended to higher orders in \cite{Iyer:1986np,Konoplya:2003ii,Matyjasek:2017psv} and combined with the usage of the Pade approximants \cite{Matyjasek:2017psv,Hatsuda:2019eoj}.
\end{enumerate}
As both methods are extensively discussed in the literature (see, for example, reviews \cite{Konoplya:2019hlu,Konoplya:2011qq}), we will not describe them in this paper, but will show that both methods are in a good agreement in the common parametric range of applicability.

The maximum of the effective potential in the eikonal regime occurs at
\begin{equation}
r_{max} = \sqrt{\frac{3}{2}} \sqrt{3 M^2+\sqrt{9 M^4+2 M^2 p^2}+p^2}.
\end{equation}
In the eikonal regime ($\ell \rightarrow \infty$) the WKB formula is correct already at the first order
\begin{equation}
\omega^2 = V_{max} - i \left(n + \frac{1}{2} \right) \sqrt{- 2 V_{max}''},
\end{equation}
where $V_{max}$ is the value of the effective potential in the peak $r = r_{max}$, and $V_{max}''$ is second derivative in the peak.
Using the expansion of the right hand side of this formula in terms of $1/\ell$ we can find the exact analytical expression for the quasinormal modes in the eikonal regime, which is relatively compact for the real oscillation frequency
$$ \frac{Re (\omega)}{2 \ell+1} = $$
\begin{equation}
\frac{\sqrt{\sqrt{2} \sqrt{3 M
   \left(\sqrt{2 a^2+9 M^2}+3
   M\right)+a^2}-4 M}}{6^{3/4} \left(M
   \left(\sqrt{2 a^2+9 M^2}+3
   M\right)+a^2\right)^{3/4}},
\end{equation}
but not for the damping rate. Therefore, a more compact expression can be obtained when expanding in terms of a small parameter $a$:

\begin{equation}\label{eikonalRe}
Re (\omega)= \frac{2 \ell+1}{6 \sqrt{3} M}- \frac{a^2 (2
   \ell+1)}{72 \left(\sqrt{3}
   M^3\right)}+\frac{11 a^4 (2 \ell+1)}{5184
   \sqrt{3} M^5}+O\left(a^6\right)
\end{equation}


\begin{equation}\label{eikonalIm}
-Im (\omega)= \frac{2 n+1}{6 \sqrt{3} M}- \frac{a^2 (2
   n+1)}{216 \left(\sqrt{3}
   M^3\right)}+\frac{a^4 (2 n+1)}{5184
   \sqrt{3} M^5}+O\left(a^6\right)
\end{equation}

Strictly speaking, the small dimensionless parameter is $a/r_{h}$, which is always smaller than unity. When $a \rightarrow 0$ the above eikonal formulas go over into their Schwarzschild forms \cite{Ferrari:1984zz}. The dependence of the quasinormal frequencies on the deformation parameter $a$ is qualitatively different for lowest and higher multipoles $\ell$. As can be seen in Tables I, II and III, both $Re (\omega)$ and $Im (\omega)$ grow when $a$ is increased up to some near extremal value of $a$. In the near extremal regime the damping rate continues growing, while the real oscillation frequency slightly decreases and approaches a constant (see Table IV.). From the analytical eikonal formulas (\ref{eikonalRe}, \ref{eikonalIm}) one can see that the first correction term to the Schwarzschild value, proportional to $a^2$ is negative, so that both real and imaginary parts of $\omega$ are smaller than their Schwazrschild limits at higher mutlipoles. This feature is missed in \cite{Saleh:2016pke,Saleh:2014uca}, because only the higher multipoles were considered there.

A general approach to finding of the eikonal quasinormal modes for spherically symmetric black hole can be found in  \cite{Churilova:2019jqx}.
This formula is also useful as the eikonal quasinormal modes of test fields are related to the parameters of the null geodesics \cite{Cardoso:2008bp,Konoplya:2017wot}.

The spectrum of the Dirac field deserves a special remark, because there are two polarizations producing the same quasinormal spectrum. Howeveer, the numerical error for the $V_{+1/2}$ potential is larger than for the  $V_{-1/2}$, especially at large $a$. Therefore, in the Table II we represent the time-domain integration data for the  $V_{-1/2}$ potential, which also produces better agreement with the $WKB$ results. In the WKB method we used the 7th order expansion with the Pade approximants chosen in such a way that $\tilde{m} =6$ (see the definition of $\tilde{m} $ and further details in \cite{Konoplya:2019hlu}). This produces the best agreement between the time-domain, WKB and accurate numerical data in the Schwarzschild limit.

\section{Greybody factors}

Calculation of grey-body factors are important, first of all for estimation of the portion of the initial quantum radiation in the vicinity of the event horizon which is reflected back to it by the potential barrier. Then the Hawking semi-classical formula can be used with the grey-body factor in order to estimate the amount of radiation which will reach the distant observer. In our case the quantization of the gravitational field implies that the system cannot be described semi-classically anymore, so that there is no much sense in using the Hawking formula, but, the meaning of the grey-body factors remains the same.

It is also essential that at least for the Schwarzschild case, radiation of test fields dominate over that of gravitons. In the semi-classical regime gravitons contribute less than two percents in the total radiation flux (see \cite{Page:1976df} and a summary on Table I in \cite{Konoplya:2019ppy}). Thus, grey-body factors of test fields are essential not only for understanding the classical scattering problem, but also for estimation of the intensity of Hawking radiation. Moreover, grey-body factors can sometimes be more influential than the temperature \cite{Konoplya:2019ppy}.

We shall consider the wave equation (\ref{wave-equation}) with the boundary conditions allowing for incoming waves from infinity. Owing to the symmetry of the scattering properties this is identical to the scattering of a wave coming from the horizon. The scattering boundary conditions for (\ref{wave-equation}) have the following form
\begin{equation}\label{BC}
\begin{array}{ccll}
    \Psi &=& e^{-i\omega r_*} + R e^{i\omega r_*},& r_* \rightarrow +\infty, \\
    \Psi &=& T e^{-i\omega r_*},& r_* \rightarrow -\infty, \\
\end{array}%
\end{equation}
where $R$ and $T$ are the reflection and transmission coefficients.
\par
The effective potential has the form of the potential barrier which monotonically decreases at both infinities, so that the WKB approach \cite{Schutz:1985zz,Iyer:1986np,Konoplya:2003ii} can be applied for finding $R$ and $T$. Since $\omega^2$ is real, the first order WKB values for $R$ and $T$ will be real \cite{Schutz:1985zz,Iyer:1986np,Konoplya:2003ii} and
\begin{equation}\label{1}
\left|T\right|^2 + \left|R\right|^2 = 1.
\end{equation}
Once the reflection coefficient is calculated, we can find the transmission coefficient for each multipole number $\ell$
\begin{equation}
\left|{\pazocal
A}_{\ell}\right|^2=1-\left|R_{\ell}\right|^2=\left|T_{\ell}\right|^2.
\end{equation}

\begin{figure}
\includegraphics[width=\linewidth]{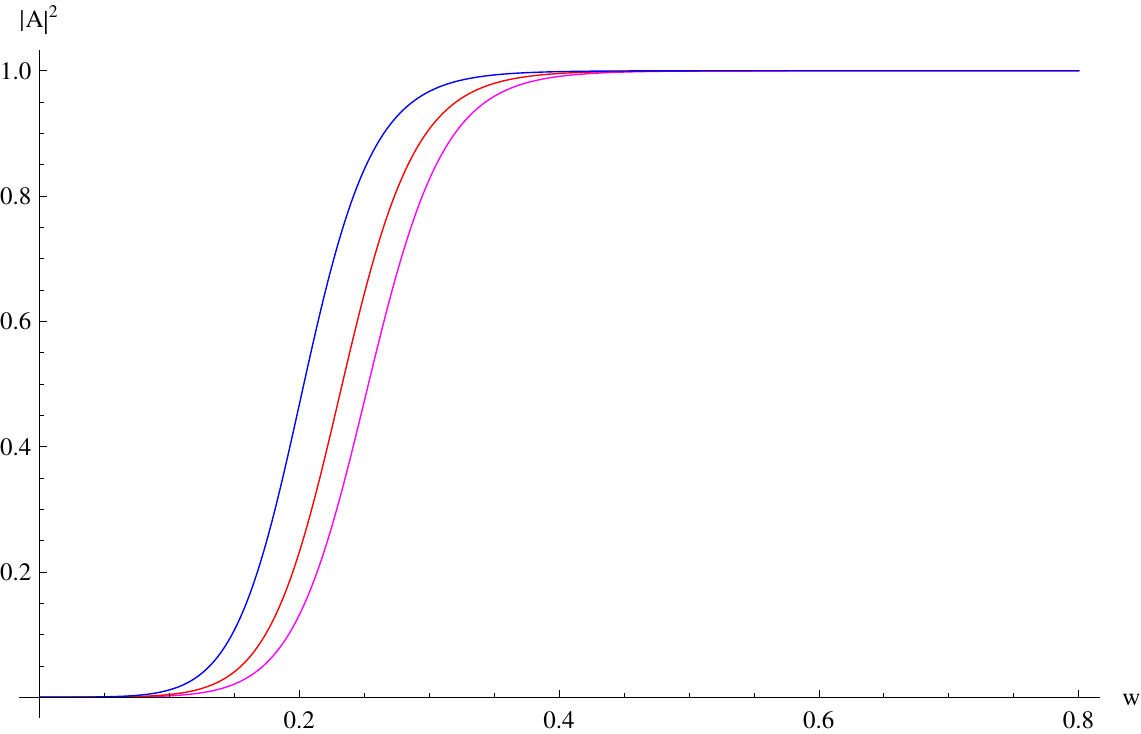}
\caption{Grey-body factors of the electromagnetic field computed with the sixth order WKB method: $M=1$, $\ell=1$, $a=0$ (the top curve on the right), $1$, $1.75$ (the bottom curve on the right) }
  \label{fig3}
\end{figure}
\begin{figure}
\includegraphics[width=\linewidth]{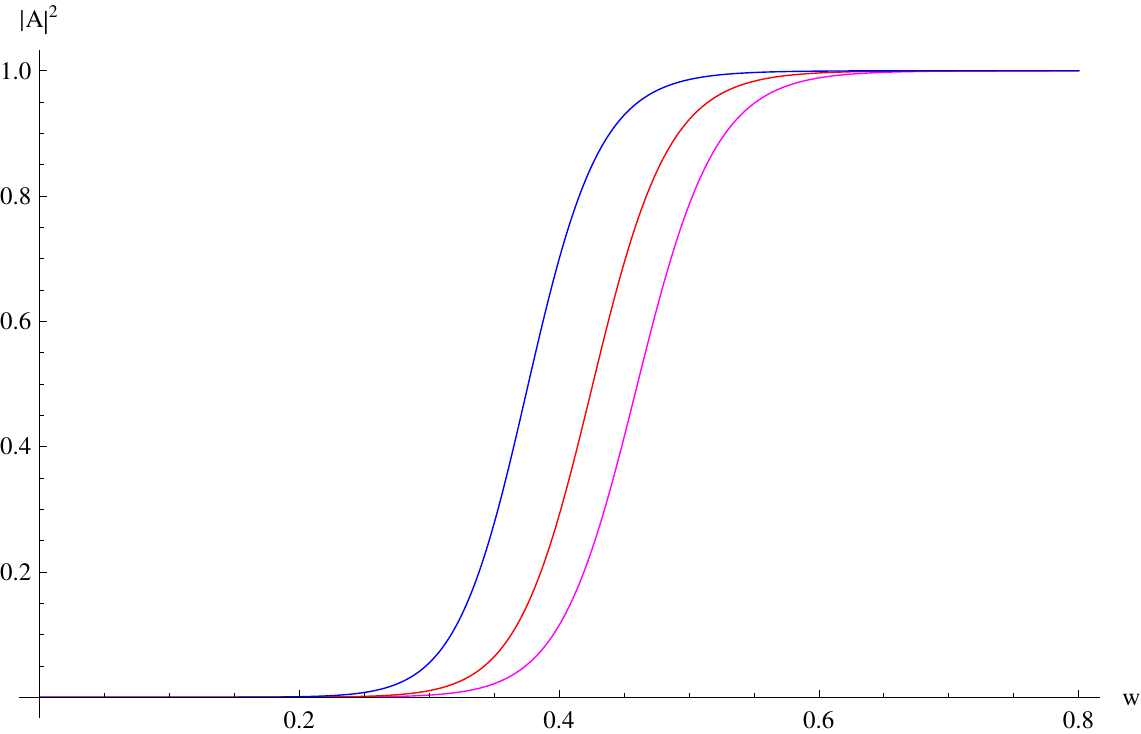}
\caption{Grey-body factors of the electromagnetic field computed with the sixth order WKB method: $M=1$, $\ell=2$, $a=0$ (the top curve on the right), $1$, $1.75$ (the bottom curve on the right) }
  \label{fig4}
\end{figure}

A number of methods for computation of the transmission and reflection coefficients exist in the literature.
For relatively accurate estimation of the transmission and reflection coefficients we used the 6th order WKB formula \cite{Konoplya:2003ii}.
The above formula does not work well when $\omega$ is very small, but fortunately, this corresponds to almost complete reflection of the waves and does not contribute considerably into the total energy emission rate. In order to study contributions of particles at very small frequencies we used the first order WKB formula which gives more accurate results in this regime. According to \cite{Schutz:1985zz,Iyer:1986np} the reflection coefficient can be expressed as follows:
\begin{equation}\label{moderate-omega-wkb}
R = (1 + e^{- 2 i \pi K})^{-\frac{1}{2}},
\end{equation}
where $K$ can be determined from the following equation:
\begin{equation}
K - i \frac{(\omega^2 - V_{max})}{\sqrt{-2 V_{max}^{\prime \prime}}} - \sum_{i=2}^{i=6} \Lambda_{i}(K) =0.
\end{equation}
Here $V_{max}$ is the maximum of the effective potential, $V_{max}^{\prime \prime}$ is the second derivative of the
effective potential in its maximum with respect to the tortoise coordinate, and $\Lambda_i$  are higher order WKB corrections which depend on up to $2i$th order derivatives of the effective potential at its maximum and $K$.

From figs. \ref{fig3},\ref{fig4} we can see that at low frequencies larger deformation $a$ corresponds to a lower grey-body factor, so that the radiation at low energies is slightly enhanced. This however should not be important for the total radiation, because this enhancement occurs in the regime of almost complete reflection.
At the same time, at intermediate frequencies the situation is opposite: larger $a$ corresponds to stronger reflection from the barrier. Finally at high frequencies quantum deformations again work for lowering the grey-body factors and enhancement of radiation (see figs. \ref{fig3},\ref{fig4}). Here we show only the first two multipoles, because contribution of higher $\ell$ into the total flux of radiation is usually negligible (see, for instance, fig. 9 in \cite{Konoplya:2019hml}). 

\section{Non-minimally coupled scalar field}

As the quantum corrected black-hole spacetime is not Ricci-flat, the evolution of non-minimally coupled fields will show even more distinction from the Schwarzschild case.
The generalization of the Klein-Gordon equation to four-dimensional spacetimes wtih non-vanishing curvature is \cite{Chernikov:1968zm}:
\begin{equation}
\Box \Phi + \frac{1}{6} R (r) \Phi =0.
\end{equation}
The corresponding effective potenatil has the form
\begin{equation}\label{scalarpotential}
V_{0}(r) = f(r)\left(\frac{\ell(\ell+1)}{r^2}+\frac{1}{r}\frac{d f(r)}{dr} + \frac{1}{6} R(r) \right),
\end{equation}
where $R(r)$ is given by (\ref{eq:R}). 

If one relies upon the time-domain data, it can be seen by comparing table I and table V that the conformal scalar field is characterized by slightly higher oscillations frequency and lower damping rate than the minimally coupled scalar field. The grey-body factors for conformal coupling differ insignificantly from those for the minimal ones adn, therefore, are not presented here.

At asymptotically late times, both the minimally coupled and conformally coupled scalar fields decay as 
\begin{equation}
|\Psi| \sim t^{- (2 \ell +3)},
\end{equation}
what coincides with the Schwarzschild limit. 

\begin{table}
\begin{tabular}{p{1cm}cccc}
\hline
$a$ & WKB (7th order, $\tilde{m} =6$) & Time-domain  \\
\hline
0.1    & $0.222623 - 0.209457 i$ &  $0.22087-0.21164 i$  \\
0.2    & $0.223720 - 0.212184 i$ &  $0.22225-0.21441 i$  \\
0.3    & $0.225647 - 0.216986 i$ &   $0.22430-0.21899 i$  \\
0.4    & $0.228572 - 0.224291 i$ &  $0.22696-0.22642 i$  \\
0.5    & $0.232779 - 0.234696 i$ &$0.22928 - 0.23749 i$  \\
0.6    & $0.238626 - 0.249205 i$  & $0.23700-0.25170 i$  \\
0.7    & $0.247631 - 0.271143 i$ &  $0.24542 - 0.27551 i$ \\
0.8    & $0.265972 - 0.307864 i$ &  $0.26073 - 0.31073 i$  \\
0.9    & $0.352638 - 0.405980 i$ & $0.29350 - 0.38222 i$  \\
\hline
\end{tabular}
\caption{The fundamental quasinormal mode of the conformally scalar field ($\ell=0$, $n=0$, $r_{h} =1$) for various values of $a$. The scalar quasinormal frequencies for $\ell=0$ cannot be easily extracted from the time domain profile at large $a$, because the asymptotic tail dominates after only a few oscillations, which is not enough for the Prony method. }\label{tab4}
\end{table}

\section{Shadows}

The radius of the photon sphere $r_{ph}$ of a spherically symmetric black hole is determined by means of the following function: (see, for example, \cite{Bisnovatyi-Kogan:2017kii} and references therein)
\begin{equation}
h^2(r) \equiv \frac{r^2}{f(r)} \,,
\label{h2 definition}
\end{equation}
as the solution to the equation
\begin{equation}
\frac{d}{dr} h^2 (r)=0\,.
\end{equation}
Then, the radius of the black-hole shadow $R_{sh}$ as seen by a distant static observer located at $r_O$ will be
\begin{equation}
R_{sh} = \frac{h(r_{ph})r_O}{h(r_O)} = \frac{r_{ph}\sqrt{f(r_O)}}{\sqrt{f(r_{ph})}} \approx  \frac{r_{ph}}{\sqrt{f(r_{ph})}}\,,
\label{shadow def}
\end{equation}
where in the last equation we have assumed that the observer is located sufficiently far away from the black hole so that $f(r_O) \approx 1$.
\begin{figure}
\includegraphics[width=0.9\linewidth]{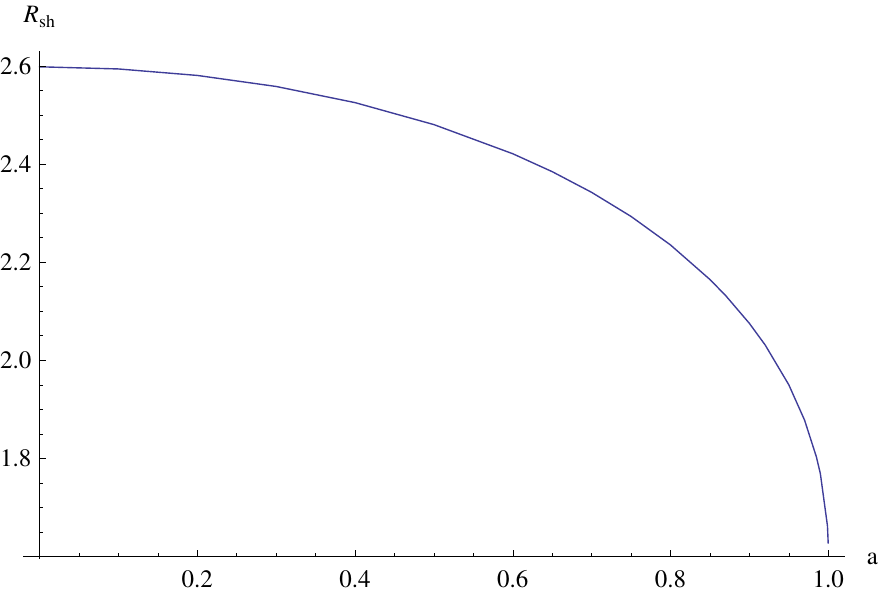}
\caption{The radius of the shadow as a function of $a$. The parameter $a$ ranges from zero to $0.99999$ on the plot; $r_{h} =1$. }
  \label{fig1}
\end{figure}
On fig. (\ref{fig1}) one can see that when the radius of the black hole is fixed, the radius of shadow is decreasing, once the quantum correction is turned on. This can be easily explained by looking at the form of the metric function $f(r)$: when the radius of the event horizon $r_{h}$ is fixed and $a$ approaches $r_{h}$, the mass of the black hole decreases and goes to zero, what should correspond to weaker gravitational attraction and smaller radius of the shadow.

\vspace{4mm}
\section{Conclusions}

Here we have considered some of the basic properties which tests the geometry of the Schwarzschild black hole with quantum correction obtained  by Kazakov and Solodukhin in \cite{Kazakov:1993ha}. This solution, known for a long time, seems to be the only one for which the problem was solved exactly and non-perturbatively, and, neglecting non-spherical deformations, the renormalizability was provided. However, this result was mainly overlooked in the literature. Here we have computed quasinormal modes of scalar, Dirac and electromagnetic fields for such black holes, obtained an analytical formula in the eikonal regime. In addition, we considered the spectrum of the conformally coupled scalar field. We also analyzed the behavior of grey-body factors in the presence of the quantum deformation, which may enhance or suppress radiation, depending on the frequency regime. The radius of the black hole shadow is shown to be decreasing, when the deformation parameter is increased.

Our paper could be extended in the following ways. The massive fields in the background of various (but not all) black holes and wormholes allow for arbitrarily long lived quasinormal modes, called quasi-resonances \cite{Ohashi:2004wr,Konoplya:2004wg,Churilova:2019qph,Konoplya:2017tvu,Zinhailo:2018ska}. It would be interesting to see whether this occurs also for the Kazakov-Solodukhin black hole. The detailed analysis of massless and massive particle motion in this spacetime would be useful as well.


\acknowledgments{
The author acknowledges  the  support  of  the  grant  19-03950S of Czech Science Foundation ($GA\check{C}R$).}

\end{document}